\documentclass[
  aps,
  pra,
  amsmath,amssymb,
  reprint, 
  showpacs
]{revtex4-1}
\usepackage{graphicx,color}
\usepackage{amsmath}
\usepackage{natbib}
\usepackage{epsfig}
\begin{document}

\title{Collective modes in two-dimensional one-component-plasma with logarithmic interaction}

\author{Sergey A. Khrapak}
\affiliation{Aix-Marseille-Universit\'{e}, CNRS, Laboratoire PIIM, UMR 7345, 13397 Marseille cedex 20, France}
\affiliation{Forschungsgruppe Komplexe Plasmen, Deutsches Zentrum f\"{u}r Luft- und Raumfahrt,
Oberpfaffenhofen, Germany}
\affiliation{Joint Institute for High Temperatures, Russian Academy of Sciences, Moscow, Russia}

\author{Boris Klumov}
\affiliation{Aix-Marseille-Universit\'{e}, CNRS, Laboratoire PIIM, UMR 7345, 13397 Marseille cedex 20, France}
\affiliation{Joint Institute for High Temperatures, Russian Academy of Sciences, Moscow, Russia}
\affiliation{L.D. Landau Institute for Theoretical Physics, Russian Academy of Sciences, Moscow, Russia}

\author{Alexey G. Khrapak}
\affiliation{Joint Institute for High Temperatures, Russian Academy of Sciences, Moscow, Russia}

\begin{abstract}
The collective modes of a familiar two-dimensional one-component-plasma with the repulsive logarithmic interaction between the particles are analysed using the quasi-crystalline approximation (QCA) combined with the molecular dynamic simulation of the equilibrium structural properties. It is found that the dispersion curves in the strongly coupled regime are virtually independent of the coupling strength. Arguments based on the excluded volume consideration for the radial distribution function allow us to derive very simple expressions for the dispersion relations, which show excellent agreement with the 
exact QCA dispersion over the entire domain of wavelengths. Comparison with the results of the conventional fluid analysis is performed and the difference is explained.     
\end{abstract}

\date{\today}
\maketitle

\section{Introduction}

The system considered in this paper represents a collection of point-like particles moving on a two-dimensional surface and interacting via a pairwise repulsive logarithmic potential of the form
\begin{equation}\label{potential}
V(r)=-\epsilon\ln(r/a),
\end{equation}
where $\epsilon$ is the energy scale and $r/a$ is the reduced distance between a pair of particles. This system is often referred to as the two-dimensional one-component-plasma (2D OCP). The interaction potential (\ref{potential}) corresponds to the solution of the 2D Poisson equation and represents the interaction between infinite charged filaments. In the conventional notation $\epsilon=Q^2$, where $Q$ is the particle charge, and $a=(\pi n)^{-1/2}$ is the 2D Wigner-Seitz radius ($n$ being the 2D number density). In addition these point charges are immersed into a rigid neutralizing background to guarantee system stability and finite values of the thermodynamical quantities.

The considered system has been employed to model vortices in thin-film semiconductors~\cite{FranzPRL_1994,MoorePRL_1999,Nattermann2000} and has some relevance in the context of the anomalous quantum Hall effect~\cite{LaughlinPRL_1983}. Exact analytic solutions for the free energy and related thermodynamic quantities in some special cases exist~\cite{Jancovici1981,Alastuey1981}. The system is characterized by ultra-soft interactions between the particles and is of interest from the fundamental point of view as an opposite limit of the celebrated hard sphere (hard disc in 2D) model. 

Thermodynamic properties of the considered 2D OCP have received considerable attention~\cite{Caillol1982,Johannesen1982, Leeuw1982,Choquard1983,Radloff1984,Herland2013,KhrapakPoP2014_OCP,KhrapakAIPAdv2015}. 
They are determined by the single parameter $\Gamma=\epsilon/T$, the so-called coupling parameter. As $\Gamma$ increases, the OCP shows a transition from a weakly coupled gaseous regime ($\Gamma\ll 1$) to a strongly coupled fluid regime ($\Gamma\gg 1$). At $\Gamma>\Gamma_{\rm m}$ (the subscript ``m'' refers to melting) the triangular lattice becomes thermodynamically favourable. Numerical simulations located this fluid-crystal phase transition at $\Gamma_{\rm m}\simeq 130 - 140$~\cite{Caillol1982,Leeuw1982,Choquard1983,Radloff1984}. In a more recent paper~\cite{MoorePRL_1999} the arguments have been presented that the crystalline state of this system is unstable at any finite temperature against proliferation of screened disclinations. This, however, remains a controversial issue~\cite{Nattermann2000} and is not the main topic of this study.

The main purpose of the present paper is to investigate the behaviour of collective modes in a strongly coupled 2D OCP fluid with the logarithmic interaction. We apply the quasi-crystalline approximation (QCA) of Hubbard and Beeby~\cite{Hubbard1969} to obtain the dispersion relations of the longitudinal and transverse modes at strong coupling. To do this we combine the QCA method with molecular dynamic (MD) simulations to obtain accurate structural properties of the system. The main results are as follows.  We demonstrate that in appropriate reduced units the dispersion relations are not very sensitive to exact structural properties, and therefore very weakly depend on the coupling strength. Moreover, based on the simplified QCA version discussed recently~\cite{SimpleQLCA}, we derive very simple analytic expressions for the dispersion curves, which demonstrate excellent agreement with the exact QCA results. We also compare the predictions of the QCA and the conventional fluid approach for the long-wavelength dispersion of the longitudinal waves and explain the difference between the results of these two approaches.    

To conclude the introductory part, we note that there is another, perhaps more familiar system, which is also referred to as the 2D OCP. In this system the interaction between the particles is of conventional Coulomb type ($\propto 1/r$), but the particle motion is restricted to a 2D surface. This system has been used as a first approximation for the description of
electron layers bound to the surface of liquid dielectrics and of inversion layers in semi-conductor physics~\cite{Baus1980,FortovBook}. It has also some relevance to colloidal and complex (dusty) plasma mono-layers in the regime of week screening~\cite{FortovBook,FortovUFN2004,FortovPR2005,IvlevBook}. Some thermodynamic properties of these two OCP fluids with Coulomb and logarithmic interactions are essentially the same. In particular, the thermal component of the internal energy exhibits the same scaling with the coupling parameter~\cite{khrapak2015nearOCP,KhrapakCPP2016}. The same scaling also takes place in the weakly screened Yukawa systems~\cite{khrapak2015nearOCP} and thus it can be considered as 2D analogy of the Rosenfeld-Tarazona scaling for soft repulsive interactions in 3D~\cite{RTScaling,RosenfeldPRE2000}. We do not elaborate on this further since the main focus of this study is on the collective excitations in 2D OCP with the logarithmic interactions.

\section{Implementation of the QCA}

The quasi-crystalline approximation was proposed in Ref.~\cite{Hubbard1969} and further detailed in Ref.~\cite{Takeno1971}. This theoretical approach can be regarded as a generalization of the phonon theory of solids. In the simplest version, the particles forming liquid are assumed stationary (i.e. like in cold amorphous solid) but the system is characterized by a liquid-like order, measured in terms of the isotropic radial distribution function (RDF) $g(r)$. The linear response of such disordered system can be approximately calculated and related to the frequencies of the collective modes~\cite{Hubbard1969}. The theory becomes exact in the special case of a cold crystalline solid~\cite{Hubbard1969}. In this sense the term ``quasi-crystalline approximation'' suggested by Takeno and Goda~\cite{Takeno1971} appears adequate and we employ it here. Comparable expressions can also be obtained from the analysis of the fourth frequency moment~\cite{deGennes1959}. In the context of plasma physics similar approach is known as the quasilocalized charge approximation (QLCA)~\cite{GoldenPoP2000}. In particular, it specifies how the presence of the neutralizing background has to be accounted for in the case of charged particle systems. In last decades the QLCA has been successively applied to describe collective modes in various strongly coupled plasma systems. In particular, this includes one-component-plasma with the Coulomb interactions~\cite{GoldenPoP2000} and complex (dusty) plasmas with Yukawa interactions~\cite{RosenbergPRE1997,KalmanPRL2000,OhtaPRL2000,
KalmanPRL2004,DonkoJPCM2008}, in both 3D and 2D situations. 

In the QCA model the dispersion relations are related to the inter-particle interaction potential $V(r)$ and the equilibrium radial distribution function $g(r)$ of strongly interacting particles. The compact QCA expressions for the longitudinal and transverse modes in neutral fluids are~\cite{Hubbard1969}
\begin{equation}
\omega_L^2=\frac{n}{m}\int\frac{\partial^2 V(r)}{\partial z^2} g(r) \left[1-\cos(kz)\right]d{\bf r},
\end{equation}
\begin{equation}
\omega_T^2=\frac{n}{m}\int\frac{\partial^2 V(r)}{\partial y^2} g(r) \left[1-\cos(kz)\right]d{\bf r}.
\end{equation} 
Here $\omega_{L}$ and $\omega_T$ are the frequencies of the longitudinal and transverse modes, $m$ is the particle mass, $k$ is the wave number, and $z=r\cos\theta$ is the direction of the propagation of the longitudinal mode. Adopted to the 2D situation (particles are confined to the $yz$ plane) and to the presence of the neutralizing background (this is detailed for the 2D and 3D OCP with Coulomb interactions in the overview of the QLCA model~\cite{GoldenPoP2000}) these expressions yield for the potential of Eq.~(\ref{potential}):   
\begin{equation}
\label{dispL}
\omega_{L}^2=\omega_0^2+\omega_0^2\int_0^{\infty}\frac{h(x)dx}{x}J_2(qx)
\end{equation}
and
\begin{equation}\label{dispT}
\omega_T^2=-\omega_0^2\int_0^{\infty}\frac{h(x)dx}{x}J_2(qx).
\end{equation}
Here $\omega_0=\sqrt{2\pi Q^2n/m}$ is the 2D plasma frequency, $x=r/a$ is the reduced distance, $h(x)=g(x)-1$ is the pair correlation function, $q=ka$ is the reduced wave number, and $J_2(x)$ is the Bessel function of the first kind. From Eqs.~(\ref{dispL}) and (\ref{dispT}) we immediately see that
\begin{displaymath}
\omega_L^2+\omega_T^2=\omega_0^2,
\end{displaymath}
which represents the two-dimensional version of the Kohn's sum rule (note that it does not apply to Coulomb interaction in 2D ~\cite{GoldenPoP2000}).

We can easily analyse two limiting cases. In the long-wavelength (small $q$) limit we use the series expansion
\begin{displaymath}
J_2(x)\simeq \frac{x^2}{8}-\frac{x^4}{96}+\frac{x^6}{3072}+{\mathcal O}(x^8)
\end{displaymath}
combined with the fist sum rules~\cite{Caillol1982}
\begin{align}
\int_0^{\infty}h(x)xdx=-\frac{1}{2}, \label{quasineutrality}\\
\int_0^{\infty}h(x)x^3dx = -\frac{1}{\Gamma}, \\
\int_0^{\infty}h(x)x^5dx = -\frac{2(4-\Gamma)}{\Gamma^2}.
\end{align}
The first sum rule, Eq.~(\ref{quasineutrality}), combined with the virial equation, immediately yields the reduced excess pressure:
\begin{equation}\label{pressure}
p_{\rm ex}\equiv\frac{P}{nT}-1=-\frac{\Gamma}{4}.
\end{equation}          
We obtain for the long-wavelength dispersion relation of the longitudinal mode: \begin{equation}\label{low_q}
\frac{\omega_L^2}{\omega_0^2}\simeq 1-\frac{q^2}{16}+\frac{q^4}{96\Gamma}-\frac{q^6(4-\Gamma)}{1536\Gamma^2}+{\mathcal O}(q^8).
\end{equation}
Note that in this normalization the first two dominant terms are independent of $\Gamma$. They coincide with the harmonic solid analysis of Ref.~\citep{Alastuey1981}.  
Higher terms do depend on $\Gamma$, but 
in the strongly coupled regime ($\Gamma\gg 1$) where QCA is applicable their role is completely negligible. This provides preliminary indication that the dispersion relations may be not very sensitive to $\Gamma$ in this normalization and we will elaborate on this further below. In the short-wavelength limit (large $q$) the longitudinal and transverse frequencies approach the common asymptote, the Einstain frequency $\omega_{\rm E}=\omega_0/\sqrt{2}$.

To go beyond these two limits, the explicit form of the radial correlation function $h(x)$ is required. We have obtained $h(x)$ using the MD numerical simulations as described in the next section.

\section{Simulations}

\begin{figure}
\includegraphics[width=5.0cm]{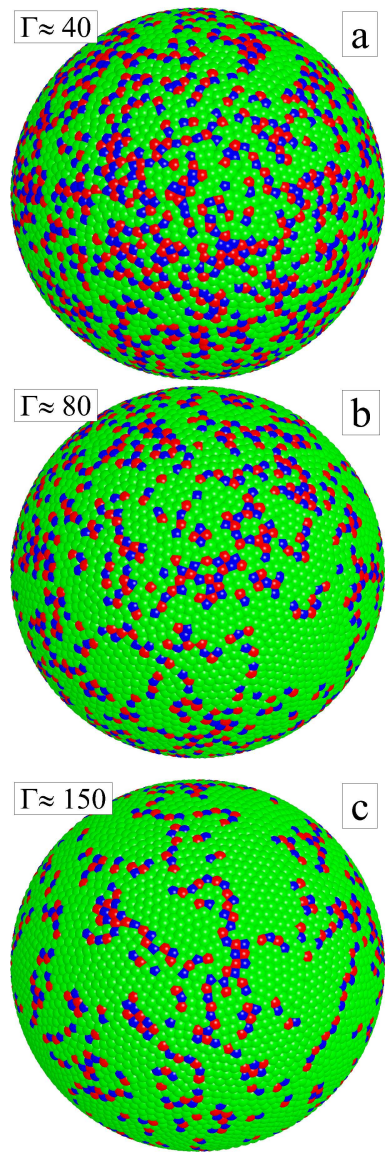}
\caption{(color online)  Two dimensional OCP with the
logarithmic interaction for different $\Gamma$ values (indicated on the plot) as modelled on a sphere. Particles are color-coded via the number of nearest neighbors ($N_{\rm nn}$)
defined from the Voronoi analysis: five-fold (blue), six-fold
(green) and seven-fold (red). By increasing the coupling strength $\Gamma$ the density of defects (five-fold and seven-fold
particles) is considerably reduced. }
\label{Fig1}
\end{figure}

Standard molecular dynamics simulations with the Verlet
velocity algorithm and Langevin thermostat have been performed (see, e.g. Ref.~\cite{Klumov2011}).  Initially,
$N = 4800$ point-like particles are randomly distributed over the unit sphere
(to eliminate the periodic boundary conditions), equilibrated (at a given $\Gamma$)
configurations are then used to calculate $g(x)$ [and hence $h(x)$] and to provide the Voronoi tessellation. Some results of the simulations are presented in Fig.~\ref{Fig1}, 
where typical configurations of particles (color-coded via the number of
nearest neighbours) are shown for the three values of $\Gamma$ ($\Gamma = 40$, 80, and 150). The ground state of the 2D
OCP with the logarithmic interaction is well known to be hexagonal, so that blue (five-fold) and red (seven-fold) particles are the topological defects. The defects
abundance, $\delta_{\rm d} = (N_{\rm 5fold} + N_{\rm 7fold})/N$, drops down as
$\Gamma$ increases. For the configurations shown in Fig.~\ref{Fig1}
these abundances are about 0.41 ($\Gamma= 40$), 0.30 ($\Gamma= 80$), and 0.22 ($\Gamma=150$). The processes of recombination and association of defects form clusters of a complicated shape, varying from 1D string-like at high $\Gamma$, to 2D-like at low $\Gamma$ (see Ref.~\cite{Klumov2011} to compare with clusters of defects in Yukawa systems). More structural details of the system at strong coupling will be reported elsewhere.

\begin{figure}
\includegraphics[width=7.5cm]{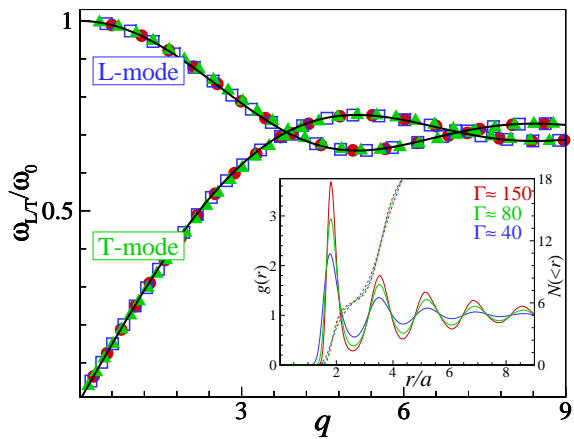}
\caption{(color online) The dispersion relations of the
longitudinal (L-mode) and transverse (T-mode) waves in
strongly coupled 2D OCP with the
logarithmic interaction. The solid black curves correspond to the
simplified QCA, Eqs.~(\ref{SQLCA1}) and ~(\ref{SQLCA2}). Symbols
correspond to conventional QCA with $g(r)$ obtained via the direct MD simulations:
red, green and blue colors corresponds to coupling parameter
$\Gamma \approx 150$, 80 and 40, respectively. The corresponding RDFs are shown in the inset. The cumulative functions $N(<r)$ of $g(r)$ are also plotted in the inset by dashed lines to evaluate the number of particles in different shells. This function clearly reveals sixfold type of symmetry in the system. }
\label{Fig2}
\end{figure}

Using the obtained RDFs $g(x)$ (plotted in the inset of Fig.~\ref{Fig2}) the dispersion curves of the longitudinal and transverse modes, within the QCA approach, have been calculated. The results are presented in Fig.~\ref{Fig2}. The symbols correspond to the results obtained using integration of Eqs.~(\ref{dispL}) and (\ref{dispT}). It is evident that in the considered regime of strong coupling, the dispersion relations are very insensitive to the exact value of $\Gamma$, although the variations in RDFs are significant. The symbols are all falling on the two distinct curves (L-mode and T-mode), no signature of any systematic deviations can be detected. The black curves correspond to the simplified version of the QCA, discussed in the next Section. At this point we just note that the agreement between these curves and the location of the symbols is excellent.

\section{Simplified QCA}

Recently, we have proposed a useful simplification of the QCA (or, equivalently, QLCA) formalism~\cite{SimpleQLCA}. In this simplified version the excluded volume arguments suggest to use a simplest toy step-wise model for $g(x)$, that is $g(x)=1$ for $x>R$ and $g(x)=0$ otherwise. Here $R$ (measured in units of $a$) characterizes the radius of an excluded sphere around each particle due to a strong (repulsive) inter-particle interaction. To estimate this quantity energy and pressure equations have been used, which are also expressed as certain integrals over $g(x)$~\cite{HansenBook}, or $h(x)$ in the present case.  For the single component Yukawa systems it  was demonstrated that the results are not particularly sensitive to whether the energy or pressure equation is used to determine $R$. The obtained dispersion relations are in very good agreement with the conventional QLCA in the long wavelength regime and correctly predict the approach to the Einstein frequency in the short-wavelength limit~\cite{SimpleQLCA}. 
  
Thus, it is tempting to check the performance of this simplified QCA in the case of 2D OCP with the logarithmic interaction considered here. In this case there is a natural way to determine $R$, by requiring the perfect screening condition (\ref{quasineutrality}) to be satisfied (in this way we also get the exact result for the excess pressure, but not for the excess energy). 
This results in $R=1$ and, hence, 
\begin{equation}\label{SQLCA1}
\frac{\omega_L^2}{\omega_0^2}= \frac{1}{2}+\frac{J_1(q)}{q},
\end{equation}
\begin{equation}\label{SQLCA2}
\frac{\omega_T^2}{\omega_0^2}= \frac{1}{2}-\frac{J_1(q)}{q}.
\end{equation}
This shows excellent agreement with the conventional (full) QCA (or QLCA) formalism in the entire domain of $q$ (see Fig.~\ref{Fig2}), much better than in the case of 3D Yukawa systems~\cite{SimpleQLCA}. The agreement is so impressive that one may think of a mathematical identity involved. This is however not the case, because the low-$q$ limit of the conventional QCA [Eq.~(\ref{low_q})] does contain some (although vanishingly small) dependence on $\Gamma$, while Eq.~(\ref{SQLCA1}) does not.  

\section{Fluid approach}

The dispersion relation of the longitudinal mode within the conventional fluid description of plasma reads
\begin{equation}\label{DR}
\omega_L^2 = \omega_0^2+k^2v_T^2\gamma \mu,
\end{equation} 
where $\gamma$ is the adiabatic index, $\mu$ is the isothermal compressibility modulus, and $v_T=\sqrt{T/m}$ is the thermal velocity. 
The adiabatic index approach unity very quickly as the coupling increases, especially for soft interactions (for an example see Fig. 3 from Ref.~\cite{SemenovPoP2015}). The isothermal compressibility modulus $\mu=(1/T)(\partial P/\partial n)_T$ is trivially related to pressure in the considered case, $\mu = 1+p_{\rm ex}$, because $p_{\rm ex}$ is density independent. We have therefore at strong coupling
\begin{equation}\label{Fluid}
\omega_L^2\simeq {\omega_0^2}-\frac{1}{8}{\omega_0^2}q^2.
\end{equation} 
The second term is immediately identified to be by a factor of two larger than the QCA approach yields, Eq.~(\ref{low_q}). There is interesting physics behind this observation, which we explain now.  

In the long-wavelength limit the QCA model, applied to neutral fluids, provides the longitudinal $c_L$ and transverse $c_T$ elastic sound velocities. These can be expressed in terms of the bulk modulus $K$ and shear modulus $G$~\cite{LL,Schofield1966}. In the 2D geometry, the corresponding expressions are: $mnc_L^2=K+G$ and $mn c_T^2=G$. 
An instantaneous sound velocity~\cite{Schofield1966} defined via $mn c_{\rm I}^2=K$ is then related to the elastic sound velocities via $c_{\rm I}^2=c_L^2-c_T^2$. This instantaneous sound velocity is in fact very close to the conventional thermodynamic sound velocity,  $c_{\rm Th}=v_{T}\sqrt{\gamma\mu}$, which can be derived from a standard fluid description~\cite{LL_Fluids}, similar to that used to obtain Eq.~(\ref{DR}). The quantitative similarity between the $c_I$ and $c_{\rm Th}$ values should be particularly pronounced for soft repulsive interactions at strong coupling. This has been directly verified for weakly screened 2D Yukawa systems near the fluid-solid phase transition~\cite{KhrapakPoP2016_1}.               
The remaining step is to account for the presence of the neutralizing background and resulting non-acoustic character of the dispersion at low $q$ for the 2D OCP. 
We introduce the instantaneous frequency $\omega_{\rm I}$ and get from the arguments above (using $\omega_L$ and $\omega_T$ obtained within QCA model)
\begin{equation}
\omega_{\rm I}^2=\omega_L^2-\omega_T^2=2\omega_L^2-\omega_0^2\simeq\omega_0^2-\frac{1}{8}\omega_0^2q^2,
\end{equation}     
which coincides with the result of the fluid approach (\ref{Fluid}). This provides us a link between the results of the QCA and the fluid (thermodynamic) approximation.

\section{Conclusion}

To summarize, we have discussed the collective modes behaviour in the two-dimensional one-component-plasma fluid with the logarithmic interaction between the particles. The dispersion relations in the strong coupling regime were obtained using the QCA (or QLCA) method coupled to the MD simulations on a sphere, to get information about the system structural properties. We have also tested the simplified QCA approach based on a toy model for the radial distribution function, which accounts for the excluded volume effects. The analytic expressions derived within this simplified QCA show excellent agreement with the actual dispersion relations within the conventional QCA, over the entire range of wave numbers. This can indicate on the particular usefulness of the simplified QCA version in two dimensions (for three dimensional Yukawa systems the performance of the simplified QCA is also good, but rather limited to long wavelengths). Finally we have discussed the results of the standard fluid approach and explained the difference it yields in comparison with the QCA.

\begin{acknowledgments}
This study was supported by the A*MIDEX grant (Nr.~ANR-11-IDEX-0001-02) funded by the French Government ``Investissements d'Avenir'' program. MD simulations were supported by the Russian Science Foundation, Project No. 14-50-00124. 
\end{acknowledgments}

\bibliographystyle{aipnum4-1}
\bibliography{References_KhrapakPoP2016}

\end{document}